\documentclass{aip-cp}

\usepackage[numbers]{natbib}
\usepackage{rotating}
\usepackage{graphicx}

\newcommand{\ltsim}{\protect\raisebox{-0.5ex}{$\:\stackrel{\textstyle <}
	{\sim}\:$}}

\begin{document}

\title{Test of the $X(3872)$ Structure in Antiproton-Nucleus Collisions}

\author[aff1,aff2]{Alexei Larionov\corref{cor1}}
\author[aff3]{Mark Strikman}
\eaddress{strikman@phys.psu.edu}
\author[aff1,aff4]{Marcus Bleicher}
\eaddress{bleicher@th.physik.uni-frankfurt.de}

\affil[aff1]{Frankfurt Institute for Advanced Studies (FIAS), 
             D-60438 Frankfurt am Main, Germany}
\affil[aff2]{National Research Centre "Kurchatov Institute", 123182 Moscow, Russia}
\affil[aff3]{Pennsylvania State University, University Park, PA 16802, USA}
\affil[aff4]{Institut f\"ur Theoretische Physik, J.W. Goethe-Universit\"at,
             D-60438 Frankfurt am Main, Germany}
\corresp[cor1]{Corresponding author: larionov@fias.uni-frankfurt.de}

\maketitle

\begin{abstract}
The present day experimental data on the $X(3872)$ decays do not allow to make clear conclusions 
on the dominating structure of this state.
We discuss here an alternative way to study its structure by means of the two-step $\bar D^*$
(or $D$) production in $\bar p A$ reactions. If this process is mediated by $X(3872)$, 
the characteristic narrow peaks of the $\bar D^*$ (or $D$) distributions in the light cone momentum 
fraction at small transverse momenta will appear. This would unambiguously signal the $D \bar D^*$ + c.c.
molecular composition of the $X(3872)$ state.
\end{abstract}

\vspace{0.5cm}



\section{Introduction}
\label{intro}

The $c \bar c$ containing $X(3872)$ state (will be denoted below as ``$X$'' for brevity) 
has been discovered by BELLE \cite{Choi:2003ue} as a peak in $\pi^+ \pi^- J/\psi$ invariant mass spectrum 
from $B^{\pm} \to K^{\pm} \pi^+ \pi^- J/\psi$ decays. The quantum numbers of $X$ are $J^{PC}=1^{++}$ 
as determined by LHCb \cite{Aaij:2013zoa} based on angular correlations in the $B^+ \to K^+ X$, 
$X \to \pi^+ \pi^- J/\psi$, $J/\psi \to \mu^+ \mu^-$ decays. 
The structure of this state is nowadays under extensive discussions.
The closeness of the $X$ mass to the two-meson threshold $D^0 \bar D^{*0}$,
$|m_{X} - m_{D^0} - m_{\bar D^{*0}}| < 1$ MeV, 
stimulated the mesonic molecular model of the $X$ state 
\cite{Tornqvist:1993ng,Tornqvist:2003na,Swanson:2004pp,Tornqvist:2004qy}
bound by pion exchange potential\footnote{We disregard the difference between the $D$ and $\bar D$
states (and similar for the $D^*$ and other charmed mesons). Thus, the overbar is dropped in many places below.}.
The size of such a molecule, i.e. the root-mean-square distance between components, can be estimated
from a binding energy $E_b$ as 
\begin{equation}
  \sqrt{\langle r^2 \rangle_{\bar D D^*}}
    \simeq \frac{1}{\sqrt{2}a} \sim  1.1-4.4~{\rm fm}~,   \label{r_rms}
\end{equation}
where $a=\sqrt{2 \mu E_b}$ is a range parameter, $\mu = m_{\bar D} m_{D^*}/(m_{\bar D}+m_{D^*})$ 
is the reduced mass. The lower limit in (\ref{r_rms}) is obtained for the charged components, 
$D^- D^{*+}$, with $E_b \simeq 8$ MeV (marginally consistent with the molecular interpretation), 
while the upper limit -- for the neutral components, 
$D^0 \bar D^{*0}$, with $E_b \simeq 0.5$ MeV.
(The recent determination of the $D^{*0}$ mass \cite{Tomaradze:2015cza} based on CLEO data results 
in even smaller binding energy $E_b < 0.2$ MeV. 
Thus, the size of the $D^0 \bar D^{*0}$ + c.c. molecule may be even larger.)
Hence, if the $X$ state has the predominant $D^0 \bar D^{*0}$ + c.c. molecular structure, it is most likely
to be a quite extended object with a size larger than the deuteron size.   
According to the recent theoretical studies \cite{Guo:2014cpb,Guo:2014taa}, 
the radiative decays $X \to \gamma J/\psi (\psi^\prime)$ are weakly sensitive to the structure 
of $X$ at large distances. The decay channel $X \to D^0 \bar D^0 \pi^0$ is more affected by wave function
at large distances. However, the actual predictions of the model calculations \cite{Guo:2014cpb} are still
quite uncertain due to low energy constants and FSI effects.
 
\begin{figure}[h]
  \centerline{\includegraphics[width=200pt]{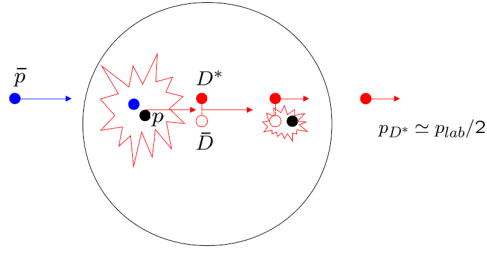}}
  \caption{\label{fig:Dstrip}A schematic view of the $D^*$ production ($\bar D$ stripping) process induced
by the antiproton annihilation on a bound proton in a nuclear target to the $X(3872)$ state assumed to be a $\bar D D^*$
molecule.}
\end{figure}
In this work we further discuss the possibility to explore the structure of $X(3872)$ by using antiproton-nucleus
reactions proposed in our recent paper \cite{Larionov:2015nea}. 
It is expected that $X$ is strongly coupled to the $\bar p p$ channel \cite{Braaten:2007sh} and, thus, 
can be produced in a $\bar p p \to X$ exclusive reaction.
In the case of a nuclear target, the produced $X$ will propagate in the nuclear residue
and possibly experience the stripping reaction on a nucleon, as illustrated in Figure~\ref{fig:Dstrip}.
Since the relative motion of the $\bar D$ and $D^*$ in a molecule is slow, the outgoing $D^*$ will propagate
in a forward direction with momentum $\sim p_{\rm lab}/2$. In terms of a light cone momentum fraction,
\begin{equation}
   \alpha=\frac{2(\omega_{D^*}+k^z)}{E_{\bar p}+m_p+p_{\rm lab}}~,         \label{alpha_Acm}
\end{equation}
this corresponds to $\alpha \simeq 1$. Here, $\omega_{D^*}(\mathbf{k})=(\mathbf{k}^2+m_{D^*}^2)^{1/2}$,
$E_{\bar p}=(p_{\rm lab}^2+m_p^2)^{1/2}$, and $z$ axis is chosen along the beam momentum.



\section{Model}
\label{model}

In order to calculate the process of Fig.~\ref{fig:Dstrip}, we have to know the two main ingredients:
the production rate $\bar p p \to X$, and the cross section of the process $X p \to D^*$.

The molecule production rate (see Eq.(\ref{Gamma_barp_to_R}) below) is proportional to the modulus squared of
the matrix element. The latter can be expressed via detailed balance as
\begin{equation}
   \overline{|M_{X;\bar p p}|^2}
   = \frac{4 \pi (2J_X+1) m_X^2 \Gamma_{X \to \bar p p}}{\sqrt{m_X^2-4m_p^2}}~,        \label{MformAver}
\end{equation}
where an overline means summation over helicity of $X$ and averaging over helicities of $\bar p$ and $p$.
The partial decay width $X \to \bar p p$ has been theoretically estimated in \cite{Braaten:2007sh} to be 
$\Gamma_{X \to \bar p p} \simeq 30$ eV.

\begin{figure}[ht]
\centerline{\includegraphics[width=300pt]{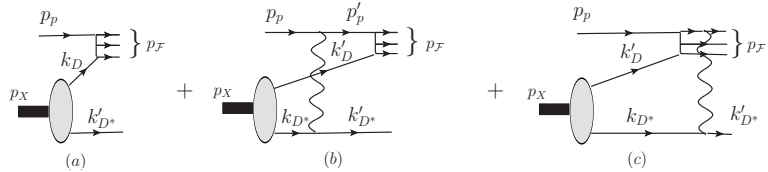}}
\caption{\label{fig:pDDbar_to_F} The amplitude for the process $X(3872)+p \to D^* + {\cal F}$ 
where ${\cal F} \equiv \{{\cal F}_1,\ldots,{\cal F}_n\}$ is an arbitrary final state in the $pD$ interaction.
Wavy lines denote the elastic scattering amplitudes. Straight lines are labelled with particle's four-momenta.
The blob represents the wave function of the molecule.}
\end{figure}
The amplitude of the $D$-stripping process with arbitrary final states is shown in Fig.~\ref{fig:pDDbar_to_F}.
We take into account the impulse approximation (IA) graph (a) and the graphs where either incoming (b) or outgoing (c) 
proton (or the most energetic forward product of the inelastic $pD$ interaction) rescatters elastically 
on the $D^*$ meson. 
The differential cross section of $D^*$ production due to the $D$ stripping from the molecule $X$ in the collision 
with a proton can be written in the molecule rest frame as
\begin{equation}
   \frac{d^3 \sigma_{p X \to D^*}}{d^3 k} =
      \sigma_{pD}^{\rm tot} {\cal I}_{pD}(-\mathbf{k}) |\psi(\mathbf{k})|^2 \kappa~,   \label{dsigma_D^*}
\end{equation}
where $\sigma_{pD}^{\rm tot}$ is the total $pD$ interaction cross section,
\begin{equation}
   {\cal I}_{pD}(\mathbf{k}) = \frac{[(E_p \omega_{D}-p_{\rm p}k^z)^2-(m_p m_{D})^2]^{1/2}}%
                                          {p_{\rm p}\omega_{D}}                       \label{calI_pD}
\end{equation}
is the Moeller flux factor (normalized to 1 for $D$ at rest), $\psi(\mathbf{k})$ is the wave function of the molecule.
$\kappa$ is a factor taking into account the screening and antiscreening corrections:
\begin{eqnarray}
   \kappa &=& 1-\sigma_{pD^*}^{\rm tot} {\cal I}_{pD^*}(\mathbf{k})  
            \int \frac{d^2 q_{t}}{(2\pi)^2} 
             \frac{\psi^*(\mathbf{k}+\mathbf{q}_{t})}{\psi^*(\mathbf{k})}
             \mbox{e}^{-(B_{pD}+B_{pD^*})\mathbf{q}_{t}^2/2}  \nonumber \\
    && +\frac{(\sigma_{pD^*}^{\rm tot} {\cal I}_{pD^*}(\mathbf{k}))^2}{4} 
        \int \frac{d^2q_{t} d^2q_{t}^\prime}{(2\pi)^4}
        \frac{\psi(\mathbf{k}+\mathbf{q}_{t}) \psi^*(\mathbf{k}+\mathbf{q}_{t}^\prime)}%
             {|\psi(\mathbf{k})|^2}
                   \mbox{e}^{-[B_{pD^*}(\mathbf{q}_{t}^2+\mathbf{q}_{t}^{\prime 2})
                  +B_{pD}(\mathbf{q}_{t}^\prime-\mathbf{q}_{t})^2]/2}~,     \label{kappa}     
\end{eqnarray}
where we used the expression for the elementary $pD$ elastic scattering amplitude
\begin{equation}
   M_{pD}(\mathbf{q}_{t})= 2 i p_{\rm p}\omega_{D} {\cal I}_{pD}(\mathbf{k}_{D})
                                 \sigma_{pD}^{\rm tot} \mbox{e}^{-B_{pD} q_{t}^2/2}~,   \label{M_pD_qt}
\end{equation}
with $\mathbf{q}_{t}$ being the transverse momentum transfer. (Expressions for for the flux factor ${\cal I}_{pD^*}(\mathbf{k})$
and for the amplitude $M_{pD^*}(\mathbf{q}_{t})$ of $pD^*$ scattering are given by Eqs.(\ref{calI_pD}),(\ref{M_pD_qt}) 
with replacement $D \to D^*$.)  

In the summation over final states ${\cal F}$ we used the unitarity relation \cite{BLP}:
\begin{equation}
    M_{fi} - M_{if}^* = \sum_{{\cal F}} \frac{d^3p_{{\cal F}_1}}{2E_{{\cal F}_1}(2\pi)^3} \cdots
                                        \frac{d^3p_{{\cal F}_n}}{2E_{{\cal F}_n}(2\pi)^3}
                                        i (2\pi)^4 \delta^{(4)}(p_{\cal F}-p_f) M_{{\cal F}f}^* M_{{\cal F}i}~,  
                                                          \label{unitarity_relation}
\end{equation}     
where '$i$' and '$f$' are the elastic scattering states of the $pD$ system.  
In the impulse approximation $\kappa=1$ which is quite accurate for small transverse momenta, $k_t \ltsim 0.1$ GeV/c.
The second (negative) term in Eq.(\ref{kappa}) is the screening correction due to the interference
of the IA amplitude (a) with the rescattering amplitudes (b) and (c) of Fig.~\ref{fig:pDDbar_to_F}.
The third (positive) term is the antiscreening correction due to the modulus squared of the sum of (b) and (c) amplitudes.

The total cross sections of $pD$ and $pD^*$ interactions are estimated as $\sigma_{pD}^{\rm tot} \simeq \sigma_{pD^*}^{\rm tot}
\simeq \sigma_{\pi^+ p}^{\rm tot}(p_{\rm lab}/2)/2 \simeq 14$ mb based on the color dipole model and comparison of the mesonic radii.
(Here, $p_{\rm lab}=7$ GeV/c is the antiproton beam momentum for the on-shell $X$ production in the $\bar p p \to X$ process.) 
The slope parameters of the $pD$ and $pD^*$ scattering are estimated as $B_{pD} \simeq B_{pD^*} \simeq B_{pK^+}$  with
$B_{pK^+} = 4$ GeV$^{-2}$ as follows from the comparison of the radii of the $D, D^*$ and $K$ mesons \cite{Larionov:2015nea}.
The total $Xp$ cross section is close to the sum of the $pD$ and $pD^*$ cross sections with a screening correction
depending on the molecule wave function. In calculations, we use $\sigma_{X  p}^{\rm tot}=26$ (23) mb for
the $D^0 \bar D^{*0}$ ($D^+ D^{*-}$) component \cite{Larionov:2015nea}.

For the molecule wave function we adopt the asymptotic solution of a Schroedinger equation at large distances,
\begin{equation}
   \psi(\mathbf{k})=\frac{a^{1/2}/\pi}{a^2+\mathbf{k}^2}~,         \label{psi_k}
\end{equation}
normalized as $\int d^3k |\psi(\mathbf{k})|^2=1$. The molecule composition is given by $86\%$ of the   
$D^0 \bar D^{*0}$ + c.c. contribution, $12\%$ of the $D^+ D^{*-}$ + c.c. contribution, and $2\%$
of the $D_s^+ D_s^{*-}$ + c.c. contribution, as it follows from the local hidden gauge calculations
\cite{Aceti:2012cb}. We neglect the small $D_s^+ D_s^{*-}$ + c.c. component in calculations.

In order to calculate the differential cross section of $D^*(D)$ production in $\bar pA$ interactions
we apply the generalized eikonal approximation \cite{Frankfurt:1996xx,Sargsian01}. This method is
based on the Feynman graph representation of the multiple scattering process and on the three assumptions:
nonrelativistic motion of nucleons in the initial and final nuclei;
no energy transfer in the multiple soft scatterings;
no longitudinal momentum transfer in elementary amplitudes. By keeping the leading order (absorptive) 
term in the scattering expansion, i.e. neglecting the product terms in the matrix element squared with
the same nucleons-scatterers in the direct and conjugated matrix elements, we obtain the 
Glauber-type expression for the differential cross section:
\begin{eqnarray}
         \alpha \frac{d^3\sigma_{\bar p A \to D^*}}{d\alpha d^2k_t} &=&
 v_{\bar p}^{-1} \int d^3 r_1\, 
  \mbox{e}^{-\sigma_{\bar p N}^{\rm tot}
                  \int\limits_{-\infty}^{z_1} dz\,\rho(\mathbf{b}_1,z)}
                 \int d^2 p_{1t}\, \frac{d^2\Gamma_{\bar p}^{1 \to X}(\mathbf{r}_1)}{d^2p_{1t}}\,
                 G_X^{p \to D^*}(\alpha,\mathbf{k}_t-\frac{\alpha}{2}\mathbf{p}_{1t}) \nonumber \\  
   && \times \int\limits_{z_1}^{\infty}dz_2\, 
  \mbox{e}^{-\sigma_{X  N}^{\rm tot}
                  \int\limits_{z_1}^{z_2} dz\,\rho(\mathbf{b}_1,z)}\,
             \rho(\mathbf{b}_1,z_2)\, 
  \mbox{e}^{-\sigma_{D^* N}^{\rm tot}
                  \int\limits_{z_2}^{\infty} dz\,\rho(\mathbf{b}_1,z)}~. 
                                                                  \label{dsigma_barpA_to_DsX}
\end{eqnarray}
Here, 
\begin{equation}
  G_X^{p \to D^*}(\alpha,\mathbf{k}_t) 
   \equiv  \omega_{D^*}\frac{d^3 \sigma_{X p \to D^*}}{d^3 k}
   = \alpha \frac{d^3 \sigma_{X p \to D^*}}{d\alpha d^2 k_t}       \label{dsigma_D^*_cov}
\end{equation}
is the invariant cross section of $D^*$ production (or $D$-stripping),
\begin{equation}
   \frac{d^2\Gamma_{\bar p}^{1 \to X}(\mathbf{r}_1)}{d^2p_{1t}} 
         = \frac{\overline{|M_{X;\bar p 1}|^2}\,v_{\bar p}}{(2\pi)^2 4 p_{\rm lab}^2 E_1}
            n_p(\mathbf{r}_1;\mathbf{p}_{1t},\Delta_{m_X}^0)     \label{Gamma_barp_to_R}
\end{equation}
is the in-medium width of $\bar p$ with respect to the production of $X$ with transverse momentum 
$\mathbf{p}_{1t}$, $v_{\bar p}=p_{\rm lab}/E_{\bar p}$ is the antiproton velocity,
\begin{equation}
   \Delta_{m_X}^0=\frac{m_p^2+E_1^2+2E_{\bar p}E_1-m_X^2}{2p_{\rm lab}}    \label{Delta_m_X}
\end{equation}
is the longitudinal momentum of the struck proton obtained from the condition that the produced
$X$ is on the mass shell, i.e. $\Delta_{m_X}^0=p_1^z,~~(p_{\bar p}+p_1)^2=m_X^2$.
The quantity $n_p(\mathbf{r}_1;\mathbf{p}_{1t},\Delta_{m_X}^0)$ in Eq.(\ref{Gamma_barp_to_R})
is the proton phase space occupation number. We apply a model 
where the local Fermi distribution is complemented with a high-momentum tail due to the short 
range proton-neutron correlations \cite{Frankfurt:1981mk}:
\begin{equation}
   n_p(\mathbf{r};\mathbf{p}) = (1-P_2) \Theta(p_F-p) 
            + \frac{\pi^2 P_2 \rho_p  |\psi_d(p)|^2 \Theta(p-p_F)}%
{\int\limits_{p_F}^\infty dp^\prime p^{\prime 2} |\psi_d(p^\prime)|^2}~,      \label{nWithTail}
\end{equation}
where $p_F(\mathbf{r})=(3\pi^2\rho_p(\mathbf{r}))^{1/3}$ is the local Fermi momentum of protons,
$\rho_p(\mathbf{r})$ is the proton density,
$P_2 \simeq 0.25$ is the proton fraction above Fermi surface,
and $\psi_d(p)$ is the deuteron wave function.

\section{Results}
\label{results}

\begin{figure}[h]
  \centerline{\includegraphics[width=300pt]{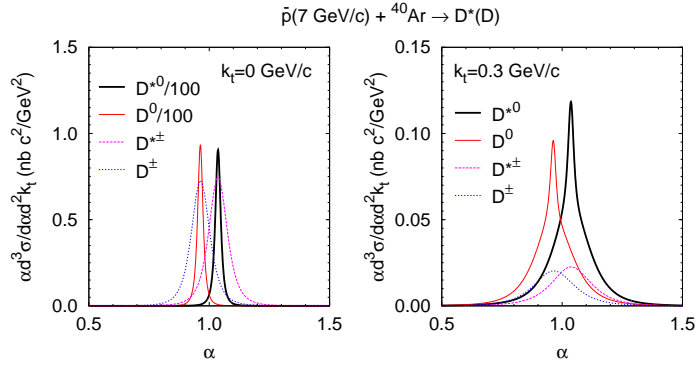}}
\caption{\label{fig:pbarA_to_Ds}  The invariant differential
cross sections of $D^{*0}$, $D^0$, $D^{*\pm}$ and $D^\pm$ production in $\bar p ^{40}$Ar collisions 
at $p_{\rm lab}=7$ GeV/c vs light cone momentum fraction $\alpha$ at $k_t=0$ (left panel) 
and $k_t=0.3$ GeV/c (right panel). For $k_t=0$, the cross sections of 
$D^{*0}$ and $D^0$ production are divided by a factor of 100.} 
\end{figure}
Figure \ref{fig:pbarA_to_Ds} shows the invariant differential cross section of $D^*$ and $D$ production
(\ref{dsigma_barpA_to_DsX}) as a function of the light cone momentum fraction $\alpha$ defined by Eq.(\ref{alpha_Acm}) 
at the two different values of the transverse momentum.
At $k_t=0$, the cross section reveals sharp peaks at $\alpha=2m_D^*/m_X=1.04$ for $D^*$ and $\alpha=2m_D/m_X=0.96$ 
for $D$. The peaks are much higher and narrower for $D^{*0}$ and $D^0$ as compared to $D^{*\pm}$ and $D^\pm$.
This is due to larger probability to find the charge neutral $D^0 \bar D^{*0}$+c.c. configuration in the molecule 
and due to its smaller binding energy.  With increasing transverse momentum the peaks gradually become smeared. 
It is, therefore, important that the transverse momentum of the outgoing $D^*$ ($D$) is
small enough, $k_t \ltsim 0.1$ GeV/c, in order the stripping signal to be visible.

\begin{figure}[h]
  \centerline{\includegraphics[width=150pt]{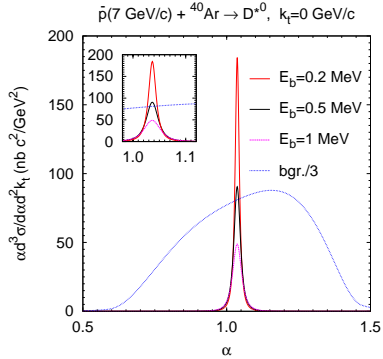}}
\caption{\label{fig:err}  The $\alpha$-dependence of $D^{*0}$ production at $k_t=0$ 
in $\bar p ^{40}$Ar collisions at $p_{\rm lab}=7$ GeV/c.
The signal cross section due to $D$-stripping from the intermediate $X$ is shown for 
the different binding energies, $E_b$, of the $D^0\bar D^{*0}$ molecule.
The background cross section is divided by a factor of 3.
The inset shows a narrower region of $\alpha$.}
\end{figure}
The major background for the $X$-mediated $D^*$ (or $D$) production is given by the direct process 
$\bar p N \to D \bar D^*$+c.c. on the bound nucleon. 
The cross section of the $\bar p p \to D^{*0} \bar D^0$ process has been estimated in \cite{Braaten:2007sh}
from dimensional counting considerations based on the measured $\bar p p \to K^{*-} K^+$ cross section.
Using the result of ref. \cite{Braaten:2007sh} as an input, we have calculated the background cross section
of $D^{*0}$ production. As one can see from Fig.~\ref{fig:err}, the background cross section is much broader
distributed in $\alpha$ than the signal, i.e. the $X$-mediated cross section.   

The binding energy of the molecule is the most crucial parameter which strongly influences the height and the 
width of the $\alpha$-distribution for the signal cross section. This is also quantified in Fig.~\ref{fig:err}, 
where the calculations are shown for the three different values of the molecule binding energy. 
We observe that such a small binding energy like $E_b \sim 0.2$ MeV \cite{Tomaradze:2015cza} leads
to an extremely sharp peak. The experimental identification of such peak would require quite high resolution
of the light cone momentum fraction, $\Delta\alpha \sim 0.01$.

\begin{figure}[h]
  \centerline{\includegraphics[width=150pt]{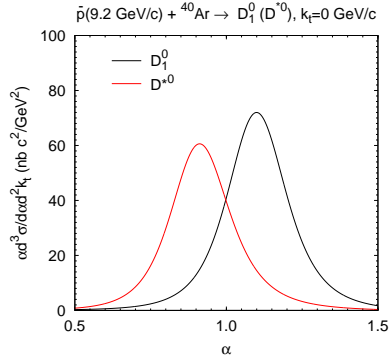}}
\caption{\label{fig:Y4360}  The $\alpha$-dependence of $D^{*0}+\bar D^{*0}$ and $D_1^0+\bar D_1^0$
production at $k_t=0$ in $\bar p ^{40}$Ar collisions at $p_{\rm lab}=9.2$ GeV/c due 
to the stripping reaction with intermediate $Y(4360)$ state.}
\end{figure}
The $X(3872)$ state is the lightest exotic $c \bar c$ state. There are several exotic states containing 
a $c \bar c$ pair which are not fit in the charmonium systematics, e.g. charge-neutral ones,
$X(3940)$, $Y(4140)$, $X(4160)$, $Y(4260)$, $Y(4360)$, and the charged ones, $Z_c(3900), Z_c(4020)$ 
(cf. \cite{Godfrey:2008nc,Brambilla:2010cs,Cleven:2015era}). The charged states are likely to be
the compact tetraquarks \cite{Cleven:2015era}. However, the neutral ones have possible molecular
structures which can also be tested in $\bar p A$ reactions in a similar way as $X(3872)$. 
In particular, the $1^{--}$ state $Y(4360)$ may be the bound state of the $D^{*0} \bar D_1^0$ + c.c. 
with a binding energy of 67 MeV \cite{Cleven:2015era}. In this case, the $\alpha$-distribution of the 
$D^{*0}$ and $D_1^0$ at $k_t=0$ due to the stripping reaction is shown 
in Fig.~\ref{fig:Y4360}. In calculations we assumed the branching ratio 
$\Gamma_{Y(4360) \to \bar p p}/\Gamma_{Y(4360)}^{\rm tot}=10^{-4}$, with the total width 
$\Gamma_{Y(4360)}^{\rm tot}=74$ MeV. Since the mass difference of $D^{*0}$ and $D_1^0$ mesons
is large, $\sim 414$ MeV, the peaks of $D^{*0}$ and $D_1^0$ distributions
in $\alpha$ are well separated. Due to the large binding energy of $Y(4360)$ state, the peaks 
are much smoother than in the case of $X(3872)$. Assuming the same shape of the $\alpha$-dependence
of the background as for $X(3872)$ such peaks would be visible as the bumps 
in the differential production cross section of $D^{*0}$ 
(at $\alpha \simeq 0.9$) and $D_1^0$ (at $\alpha \simeq 1.1$) at $k_t=0$.

\section{Conclusions and outlook}
\label{concl}

We have demonstrated that the possible $D \bar D^*$+c.c. molecular structure of $X(3872)$
manifests itself in the sharp peaks of exclusive $D^*$ or $D$ production at $\alpha \simeq 1$
for small transverse momenta. These peaks are caused by the stripping reaction of one of the 
molecular components of $X$ on the nucleon and are well visible on the smooth background 
due to the direct production of charmed mesons in $\bar p N$ collision.

Other possible structures of $X$, e.g. charmonium, tetraquark or $c \bar c$-gluon hybrid,
should produce more flat $\alpha$-distributions of $D^*$ and $D$ due to more violent 
production mechanisms in $XN$ collisions. Most likely, in these cases the charmed mesons
will be uniformly distributed in the available phase space volume in the $XN$ 
center-of-mass frame. Thus, the proposed observable, i.e. the light cone momentum fraction
distributions of $D^*$ and $D$ at small $k_t$, should be very sensitive to the hypothetical
molecular structure of $X$ state and, probably, of the other exotic $c \bar c$ candidates.
Similar processes can be considered to investigate the possible molecular structures of other
hadrons. For example, the assumed $K \bar K$ molecule composition of $a_0(980)$ and $f_0(980)$ 
mesons could be tested in a two-step process 
$\gamma(\pi) N \to f N~,~~f N \to \bar K (K) + \mbox{anything}$ 
$(f \equiv a_0(980), f_0(980))$.

The experimental studies of such processes are possible at PANDA, J-PARC, JLab and COMPASS.

\section{ACKNOWLEDGMENTS}
\label{Ack}
The work of AL has been financially supported by HIC for FAIR within the framework of the LOEWE program.

\nocite{*}
\bibliographystyle{aipnum-cp}%
\bibliography{pbarX3872_HS2015}

\begin{thebibliography}{20}%
\makeatletter
\providecommand \@ifxundefined [1]{%
 \@ifx{#1\undefined}
}%
\providecommand \@ifnum [1]{%
 \ifnum #1\expandafter \@firstoftwo
 \else \expandafter \@secondoftwo
 \fi
}%
\providecommand \@ifx [1]{%
 \ifx #1\expandafter \@firstoftwo
 \else \expandafter \@secondoftwo
 \fi
}%
\providecommand \natexlab [1]{#1}%
\providecommand \enquote  [1]{``#1''}%
\providecommand \bibnamefont  [1]{#1}%
\providecommand \bibfnamefont [1]{#1}%
\providecommand \citenamefont [1]{#1}%
\providecommand \href@noop [0]{\@secondoftwo}%
\providecommand \href [0]{\begingroup \@sanitize@url \@href}%
\providecommand \@href[1]{\@@startlink{#1}\@@href}%
\providecommand \@@href[1]{\endgroup#1\@@endlink}%
\providecommand \@sanitize@url [0]{\catcode `\$12\catcode `\&12\catcode
  `\#12\catcode `\^12\catcode `\_12\catcode `\%12\relax}%
\providecommand \@@startlink[1]{}%
\providecommand \@@endlink[0]{}%
\providecommand \url  [0]{\begingroup\@sanitize@url \@url }%
\providecommand \@url [1]{\endgroup\@href {#1}{\urlprefix }}%
\providecommand \urlprefix  [0]{URL }%
\providecommand \Eprint [0]{\href }%
\providecommand \doibase [0]{http://dx.doi.org/}%
\providecommand \selectlanguage [0]{\@gobble}%
\providecommand \bibinfo  [0]{\@secondoftwo}%
\providecommand \bibfield  [0]{\@secondoftwo}%
\providecommand \translation [1]{[#1]}%
\providecommand \BibitemOpen [0]{}%
\providecommand \bibitemStop [0]{}%
\providecommand \bibitemNoStop [0]{.\EOS\space}%
\providecommand \EOS [0]{\spacefactor3000\relax}%
\providecommand \BibitemShut  [1]{\csname bibitem#1\endcsname}%
\let\auto@bib@innerbib\@empty
\bibitem [{\citenamefont {Choi}\ \emph {et~al.}(2003)\citenamefont {Choi} \emph
  {et~al.}}]{Choi:2003ue}%
  \BibitemOpen
  \bibfield  {author} {\bibinfo {author} {\bibfnamefont {S.~K.}\ \bibnamefont
  {Choi}} \emph {et~al.} (\bibinfo {collaboration} {Belle Collaboration}),\
  }\href@noop {} {\bibfield  {journal} {\bibinfo  {journal} {Phys. Rev. Lett.}\
  }\textbf {\bibinfo {volume} {91}},\ p.\ \bibinfo {pages} {262001} (\bibinfo
  {year} {2003})}\BibitemShut {NoStop}%
\bibitem [{\citenamefont {Aaij}\ \emph {et~al.}(2013)\citenamefont {Aaij} \emph
  {et~al.}}]{Aaij:2013zoa}%
  \BibitemOpen
  \bibfield  {author} {\bibinfo {author} {\bibfnamefont {R.}~\bibnamefont
  {Aaij}} \emph {et~al.} (\bibinfo {collaboration} {LHCb}),\ }\href {\doibase
  10.1103/PhysRevLett.110.222001} {\bibfield  {journal} {\bibinfo  {journal}
  {Phys. Rev. Lett.}\ }\textbf {\bibinfo {volume} {110}},\ p.\ \bibinfo {pages}
  {222001} (\bibinfo {year} {2013})}\BibitemShut {NoStop}%
\bibitem [{\citenamefont {Tornqvist}(1994)}]{Tornqvist:1993ng}%
  \BibitemOpen
  \bibfield  {author} {\bibinfo {author} {\bibfnamefont {N.~A.}\ \bibnamefont
  {Tornqvist}},\ }\href@noop {} {\bibfield  {journal} {\bibinfo  {journal} {Z.
  Phys.}\ }\textbf {\bibinfo {volume} {C61}},\ \unskip\ \bibinfo {pages}
  {525--537} (\bibinfo {year} {1994})}\BibitemShut {NoStop}%
\bibitem [{\citenamefont {Tornqvist}(2003)}]{Tornqvist:2003na}%
  \BibitemOpen
  \bibfield  {author} {\bibinfo {author} {\bibfnamefont {N.~A.}\ \bibnamefont
  {Tornqvist}},\ }\href@noop {} {\  (\bibinfo {year} {2003})},\ \Eprint
  {http://arxiv.org/abs/hep-ph/0308277} {arXiv:hep-ph/0308277 [hep-ph]}
  \BibitemShut {NoStop}%
\bibitem [{\citenamefont {Swanson}(2004)}]{Swanson:2004pp}%
  \BibitemOpen
  \bibfield  {author} {\bibinfo {author} {\bibfnamefont {E.~S.}\ \bibnamefont
  {Swanson}},\ }\href@noop {} {\bibfield  {journal} {\bibinfo  {journal} {Phys.
  Lett. B}\ }\textbf {\bibinfo {volume} {598}},\ \unskip\ \bibinfo {pages}
  {197--202} (\bibinfo {year} {2004})}\BibitemShut {NoStop}%
\bibitem [{\citenamefont {Tornqvist}(2004)}]{Tornqvist:2004qy}%
  \BibitemOpen
  \bibfield  {author} {\bibinfo {author} {\bibfnamefont {N.~A.}\ \bibnamefont
  {Tornqvist}},\ }\href@noop {} {\bibfield  {journal} {\bibinfo  {journal}
  {Phys. Lett.}\ }\textbf {\bibinfo {volume} {B590}},\ \unskip\ \bibinfo
  {pages} {209--215} (\bibinfo {year} {2004})}\BibitemShut {NoStop}%
\bibitem [{\citenamefont {Tomaradze}\ \emph {et~al.}(2015)\citenamefont
  {Tomaradze}, \citenamefont {Dobbs}, \citenamefont {Xiao},\ and\ \citenamefont
  {Seth}}]{Tomaradze:2015cza}%
  \BibitemOpen
  \bibfield  {author} {\bibinfo {author} {\bibfnamefont {A.}~\bibnamefont
  {Tomaradze}}, \bibinfo {author} {\bibfnamefont {S.}~\bibnamefont {Dobbs}},
  \bibinfo {author} {\bibfnamefont {T.}~\bibnamefont {Xiao}}, \ and\ \bibinfo
  {author} {\bibfnamefont {K.~K.}\ \bibnamefont {Seth}},\ }\href@noop {}
  {\bibfield  {journal} {\bibinfo  {journal} {Phys. Rev.}\ }\textbf {\bibinfo
  {volume} {D91}},\ p.\ \bibinfo {pages} {011102} (\bibinfo {year}
  {2015})}\BibitemShut {NoStop}%
\bibitem [{\citenamefont {Guo}\ \emph {et~al.}(2015{\natexlab{a}})\citenamefont
  {Guo}, \citenamefont {Hidalgo-Duque}, \citenamefont {Nieves}, \citenamefont
  {Ozpineci},\ and\ \citenamefont {Valderrama}}]{Guo:2014cpb}%
  \BibitemOpen
  \bibfield  {author} {\bibinfo {author} {\bibfnamefont {F.~K.}\ \bibnamefont
  {Guo}}, \bibinfo {author} {\bibfnamefont {C.}~\bibnamefont {Hidalgo-Duque}},
  \bibinfo {author} {\bibfnamefont {J.}~\bibnamefont {Nieves}}, \bibinfo
  {author} {\bibfnamefont {A.}~\bibnamefont {Ozpineci}}, \ and\ \bibinfo
  {author} {\bibfnamefont {M.~P.}\ \bibnamefont {Valderrama}},\ }\bibfield
  {booktitle} {\emph {\bibinfo {booktitle} {{Proceedings, 5th International
  Conference on Exotic Atoms and Related Topics (EXA2014)}}},\ }\href@noop {}
  {\bibfield  {journal} {\bibinfo  {journal} {Hyperfine Interact.}\ }\textbf
  {\bibinfo {volume} {234}},\ \unskip\ \bibinfo {pages} {125--132} (\bibinfo
  {year} {2015}{\natexlab{a}})}\BibitemShut {NoStop}%
\bibitem [{\citenamefont {Guo}\ \emph {et~al.}(2015{\natexlab{b}})\citenamefont
  {Guo}, \citenamefont {Hanhart}, \citenamefont {Kalashnikova}, \citenamefont
  {Meißner},\ and\ \citenamefont {Nefediev}}]{Guo:2014taa}%
  \BibitemOpen
  \bibfield  {author} {\bibinfo {author} {\bibfnamefont {F.-K.}\ \bibnamefont
  {Guo}}, \bibinfo {author} {\bibfnamefont {C.}~\bibnamefont {Hanhart}},
  \bibinfo {author} {\bibfnamefont {{\relax Yu}.~S.}\ \bibnamefont
  {Kalashnikova}}, \bibinfo {author} {\bibfnamefont {U.-G.}\ \bibnamefont
  {Meißner}}, \ and\ \bibinfo {author} {\bibfnamefont {A.~V.}\ \bibnamefont
  {Nefediev}},\ }\href@noop {} {\bibfield  {journal} {\bibinfo  {journal}
  {Phys. Lett.}\ }\textbf {\bibinfo {volume} {B742}},\ \unskip\ \bibinfo
  {pages} {394--398} (\bibinfo {year} {2015}{\natexlab{b}})}\BibitemShut
  {NoStop}%
\bibitem [{\citenamefont {Larionov}, \citenamefont {Strikman},\ and\
  \citenamefont {Bleicher}(2015)}]{Larionov:2015nea}%
  \BibitemOpen
  \bibfield  {author} {\bibinfo {author} {\bibfnamefont {A.~B.}\ \bibnamefont
  {Larionov}}, \bibinfo {author} {\bibfnamefont {M.}~\bibnamefont {Strikman}},
  \ and\ \bibinfo {author} {\bibfnamefont {M.}~\bibnamefont {Bleicher}},\
  }\href@noop {} {\bibfield  {journal} {\bibinfo  {journal} {Phys. Lett.}\
  }\textbf {\bibinfo {volume} {B749}},\ \unskip\ \bibinfo {pages} {35--43}
  (\bibinfo {year} {2015})}\BibitemShut {NoStop}%
\bibitem [{\citenamefont {Braaten}(2008)}]{Braaten:2007sh}%
  \BibitemOpen
  \bibfield  {author} {\bibinfo {author} {\bibfnamefont {E.}~\bibnamefont
  {Braaten}},\ }\href@noop {} {\bibfield  {journal} {\bibinfo  {journal} {Phys.
  Rev. D}\ }\textbf {\bibinfo {volume} {77}},\ p.\ \bibinfo {pages} {034019}
  (\bibinfo {year} {2008})}\BibitemShut {NoStop}%
\bibitem [{\citenamefont {Berestetskii}, \citenamefont {Lifshitz},\ and\
  \citenamefont {Pitaevskii}(1971)}]{BLP}%
  \BibitemOpen
  \bibfield  {author} {\bibinfo {author} {\bibfnamefont {V.~B.}\ \bibnamefont
  {Berestetskii}}, \bibinfo {author} {\bibfnamefont {E.~M.}\ \bibnamefont
  {Lifshitz}}, \ and\ \bibinfo {author} {\bibfnamefont {L.~P.}\ \bibnamefont
  {Pitaevskii}},\ }\href@noop {} {\emph {\bibinfo {title} {Relativistic Quantum
  Theory}}}\ (\bibinfo  {publisher} {Pergamon Press},\ \bibinfo {year}
  {1971})\BibitemShut {NoStop}%
\bibitem [{\citenamefont {Aceti}, \citenamefont {Molina},\ and\ \citenamefont
  {Oset}(2012)}]{Aceti:2012cb}%
  \BibitemOpen
  \bibfield  {author} {\bibinfo {author} {\bibfnamefont {F.}~\bibnamefont
  {Aceti}}, \bibinfo {author} {\bibfnamefont {R.}~\bibnamefont {Molina}}, \
  and\ \bibinfo {author} {\bibfnamefont {E.}~\bibnamefont {Oset}},\ }\href@noop
  {} {\bibfield  {journal} {\bibinfo  {journal} {Phys. Rev. D}\ }\textbf
  {\bibinfo {volume} {86}},\ p.\ \bibinfo {pages} {113007} (\bibinfo {year}
  {2012})}\BibitemShut {NoStop}%
\bibitem [{\citenamefont {Frankfurt}, \citenamefont {Sargsian},\ and\
  \citenamefont {Strikman}(1997)}]{Frankfurt:1996xx}%
  \BibitemOpen
  \bibfield  {author} {\bibinfo {author} {\bibfnamefont {L.~L.}\ \bibnamefont
  {Frankfurt}}, \bibinfo {author} {\bibfnamefont {M.~M.}\ \bibnamefont
  {Sargsian}}, \ and\ \bibinfo {author} {\bibfnamefont {M.~I.}\ \bibnamefont
  {Strikman}},\ }\href@noop {} {\bibfield  {journal} {\bibinfo  {journal}
  {Phys. Rev. C}\ }\textbf {\bibinfo {volume} {56}},\ \unskip\ \bibinfo {pages}
  {1124--1137} (\bibinfo {year} {1997})}\BibitemShut {NoStop}%
\bibitem [{\citenamefont {Sargsian}(2001)}]{Sargsian01}%
  \BibitemOpen
  \bibfield  {author} {\bibinfo {author} {\bibfnamefont {M.~M.}\ \bibnamefont
  {Sargsian}},\ }\href@noop {} {\bibfield  {journal} {\bibinfo  {journal} {Int.
  J. of Mod. Phys. E}\ }\textbf {\bibinfo {volume} {10}},\ \unskip\ \bibinfo
  {pages} {405--457} (\bibinfo {year} {2001})}\BibitemShut {NoStop}%
\bibitem [{\citenamefont {Frankfurt}\ and\ \citenamefont
  {Strikman}(1981)}]{Frankfurt:1981mk}%
  \BibitemOpen
  \bibfield  {author} {\bibinfo {author} {\bibfnamefont {L.}~\bibnamefont
  {Frankfurt}}\ and\ \bibinfo {author} {\bibfnamefont {M.}~\bibnamefont
  {Strikman}},\ }\href@noop {} {\bibfield  {journal} {\bibinfo  {journal}
  {Phys. Rep.}\ }\textbf {\bibinfo {volume} {76}},\ \unskip\ \bibinfo {pages}
  {215--347} (\bibinfo {year} {1981})}\BibitemShut {NoStop}%
\bibitem [{\citenamefont {Godfrey}\ and\ \citenamefont
  {Olsen}(2008)}]{Godfrey:2008nc}%
  \BibitemOpen
  \bibfield  {author} {\bibinfo {author} {\bibfnamefont {S.}~\bibnamefont
  {Godfrey}}\ and\ \bibinfo {author} {\bibfnamefont {S.~L.}\ \bibnamefont
  {Olsen}},\ }\href@noop {} {\bibfield  {journal} {\bibinfo  {journal} {Ann.
  Rev. Nucl. Part. Sci.}\ }\textbf {\bibinfo {volume} {58}},\ \unskip\ \bibinfo
  {pages} {51--73} (\bibinfo {year} {2008})}\BibitemShut {NoStop}%
\bibitem [{\citenamefont {Brambilla}\ \emph {et~al.}(2011)\citenamefont
  {Brambilla}, \citenamefont {Eidelman}, \citenamefont {Heltsley},
  \citenamefont {Vogt}, \citenamefont {Bodwin} \emph
  {et~al.}}]{Brambilla:2010cs}%
  \BibitemOpen
  \bibfield  {author} {\bibinfo {author} {\bibfnamefont {N.}~\bibnamefont
  {Brambilla}}, \bibinfo {author} {\bibfnamefont {S.}~\bibnamefont {Eidelman}},
  \bibinfo {author} {\bibfnamefont {B.~K.}\ \bibnamefont {Heltsley}}, \bibinfo
  {author} {\bibfnamefont {R.}~\bibnamefont {Vogt}}, \bibinfo {author}
  {\bibfnamefont {G.~T.}\ \bibnamefont {Bodwin}},  \emph {et~al.},\ }\href@noop
  {} {\bibfield  {journal} {\bibinfo  {journal} {Eur. Phys. J. C}\ }\textbf
  {\bibinfo {volume} {71}},\ p.\ \bibinfo {pages} {1534} (\bibinfo {year}
  {2011})}\BibitemShut {NoStop}%
\bibitem [{\citenamefont {Cleven}\ \emph {et~al.}(2015)\citenamefont {Cleven},
  \citenamefont {Guo}, \citenamefont {Hanhart}, \citenamefont {Wang},\ and\
  \citenamefont {Zhao}}]{Cleven:2015era}%
  \BibitemOpen
  \bibfield  {author} {\bibinfo {author} {\bibfnamefont {M.}~\bibnamefont
  {Cleven}}, \bibinfo {author} {\bibfnamefont {F.-K.}\ \bibnamefont {Guo}},
  \bibinfo {author} {\bibfnamefont {C.}~\bibnamefont {Hanhart}}, \bibinfo
  {author} {\bibfnamefont {Q.}~\bibnamefont {Wang}}, \ and\ \bibinfo {author}
  {\bibfnamefont {Q.}~\bibnamefont {Zhao}},\ }\href@noop {} {\bibfield
  {journal} {\bibinfo  {journal} {Phys. Rev.}\ }\textbf {\bibinfo {volume}
  {D92}},\ p.\ \bibinfo {pages} {014005} (\bibinfo {year} {2015})}\BibitemShut
  {NoStop}%
\bibitem [{\citenamefont {Olive}\ \emph {et~al.}(2014)\citenamefont {Olive}
  \emph {et~al.}}]{Agashe:2014kda}%
  \BibitemOpen
  \bibfield  {author} {\bibinfo {author} {\bibfnamefont {K.~A.}\ \bibnamefont
  {Olive}} \emph {et~al.} (\bibinfo {collaboration} {Particle Data Group}),\
  }\href@noop {} {\bibfield  {journal} {\bibinfo  {journal} {Chin. Phys. C}\
  }\textbf {\bibinfo {volume} {38}},\ p.\ \bibinfo {pages} {090001} (\bibinfo
  {year} {2014})}\BibitemShut {NoStop}%
\end{thebibliography}%

\end{document}